\documentclass[11pt,a4paper]{article}

\usepackage[utf8]{inputenc}
\usepackage[T1]{fontenc}
\usepackage{lmodern}
\usepackage[margin=1in]{geometry}
\usepackage{microtype}
\usepackage{booktabs}
\usepackage{tabularx}
\usepackage{array}
\usepackage{multirow}
\usepackage{longtable}
\usepackage{enumitem}
\usepackage{amsmath,amssymb}
\usepackage{graphicx}
\usepackage{xcolor}
\usepackage{tikz}
\usetikzlibrary{arrows.meta,positioning,shapes.geometric,fit,calc}
\usepackage[hidelinks,breaklinks=true]{hyperref}
\usepackage{xurl}
\usepackage[numbers,sort&compress]{natbib}
\usepackage{caption}

\newcolumntype{L}[1]{>{\raggedright\arraybackslash}p{#1}}
\newcommand{\yes}{\textbf{\textcolor{black!80}{$\bullet$}}}
\newcommand{\prt}{\textcolor{black!50}{$\circ$}}
\newcommand{\no}{\textcolor{black!25}{--}}

\title{\textbf{TRACE: A Threat Modelling Methodology for Distributed,\\
Cloud-First, and Decentralized Organisations}}

\author{
  Stefan Beyer\\
  \small Oak Security\\
  \small \texttt{stefan@oaksecurity.io}
}

\date{\today}

\begin{document}
\maketitle

\begin{abstract}
\noindent
Established threat modelling methodologies (STRIDE, PASTA, Trike, OCTAVE,
LINDDUN, attack trees, and adversary-behaviour catalogues such as MITRE
ATT\&CK) were designed for software products and enterprises with a
discernible security perimeter, a single owning organisation, and a clean
separation between technical and operational risk. Modern organisations
increasingly violate all three assumptions. They run across cloud and SaaS
control planes they do not own, distribute privileged authority across
founders, contractors, vendors, signers, committees, and automation, and
expose value through human approval ceremonies and supply-chain edges rather
than through a network boundary. In such settings the dominant failure modes
are authorised-but-malicious actors, collusion across nominally independent
parties, control-plane and CI/CD compromise, and operational mishandling of
high-value actions. Existing methods either omit these categories or place
them out of scope.
We present TRACE, a threat modelling methodology that treats threat actors,
roles, assets, critical invariants, and trust/authority edges as first-class,
evidence-linked modelling objects, and that explicitly spans three layers:
protocols, systems, and organisations. We first give a structured comparison
of nine widely used threat modelling and adversary-analysis frameworks across
ten dimensions, then characterise where each falls short in
distributed, cloud-first, zero-trust environments. We then specify TRACE: its
core model, its three application pillars, its sequential gated workflow, and
its evidence-and-traceability discipline for human-AI co-working, in which
language models accelerate extraction and coverage while senior reviewers
retain judgement over invariants, severity, and collusion. TRACE was developed
through Web3 security practice, where high-value assets, distributed authority,
and human operations are tightly coupled, but it is stack-agnostic. We discuss
its relationship to zero trust architecture and accountable Byzantine
consensus, its limitations, and open questions around empirical validation.
\end{abstract}


\section{Introduction}

Threat modelling is the discipline of reasoning about what a system must
protect, who can affect it, and how it could fail, before failure is
demonstrated empirically. Its value is highest early, when design and
operational assumptions are still cheap to change~\cite{shostak2014threat}.
Over three decades the field has produced a large toolbox: STRIDE for structured
threat elicitation~\cite{kohnfelder1999threats,microsoft-stride}, attack trees
for goal-oriented decomposition~\cite{schneier1999attack}, PASTA for
risk- and business-centric analysis~\cite{ucedavelez2015risk}, OCTAVE for
organisational risk~\cite{alberts2002octave,caralli2007allegro}, LINDDUN for
privacy~\cite{deng2011linddun}, and adversary-behaviour knowledge bases such as
the Cyber Kill Chain~\cite{hutchins2011killchain} and MITRE
ATT\&CK~\cite{strom2018attack}. Surveys by the Software Engineering Institute
catalogue these and others~\cite{shevchenko2018summary}.

Almost all of these methods carry, explicitly or implicitly, three assumptions
inherited from the era in which they matured:

\begin{enumerate}[leftmargin=1.4em,itemsep=2pt]
  \item \textbf{A perimeter.} There is an inside and an outside; the principal
  adversary is external; trust can be granted at a boundary.
  \item \textbf{A single owning organisation.} One enterprise owns the assets,
  the infrastructure, and the people who operate it; risk decomposes cleanly
  into ``the system'' and ``the company.''
  \item \textbf{Authorisation equals safety.} If every permission check passes,
  the system is behaving correctly; the threat is the actor who is \emph{not}
  authorised.
\end{enumerate}

Modern operating environments break all three. Organisations run on cloud IaaS,
SaaS, and identity providers they do not own and cannot fully inspect; the NIST
zero trust architecture guidance~\cite{nist800207} and the CISA maturity
model~\cite{cisa-ztmm} exist because the perimeter no longer
describes reality. Privileged authority is fragmented across founders,
remote employees, contractors, vendors, managed-service providers, multisig
signers, governance committees, and automation pipelines that often span
several legal entities. The most damaging recent incidents are not
unauthorised-access bugs but \emph{authorised-path} failures: a compromised
build pipeline shipping a signed update~\cite{mehdi2021solarwinds}, a signer
deceived into approving a malicious transaction through a tampered
interface~\cite{fbi2025bybit}, or a privileged role acting within its formal
permissions. Insider and supply-chain vectors now account for a substantial
and growing share of breaches~\cite{verizon2024dbir}, and in high-value
decentralized systems, operational and human-factor compromises dominate over
pure protocol exploits~\cite{chainalysis2024}.

This paper makes three contributions:

\begin{enumerate}[leftmargin=1.4em,itemsep=2pt]
  \item A \textbf{structured comparison} (Section~\ref{sec:survey}) of nine
  widely used threat modelling and adversary-analysis frameworks across ten
  dimensions, distinguishing what each does well from what it structurally
  omits.
  \item A \textbf{gap analysis} (Section~\ref{sec:gaps}) that locates these
  omissions in the distributed, cloud-first, zero-trust setting:
  invariants, distributed authority, collusion, control-plane and supply-chain
  edges, and human/operational failure.
  \item The \textbf{TRACE methodology} (Sections~\ref{sec:trace}--\ref{sec:ai}):
  its core model, three application pillars, sequential gated workflow, and an
  evidence-and-traceability discipline that makes human-AI co-working safe and
  reviewable.
\end{enumerate}

TRACE does not replace STRIDE, attack trees, or zero trust implementation work;
it composes them. It reuses STRIDE for elicitation and attack trees for depth,
but it changes the \emph{primitives} of the model so that invariants,
distributed authority, and collusion surfaces become objects the analyst is
forced to name and connect to evidence.

\section{Background: A Comparative Survey of Threat Modelling Frameworks}
\label{sec:survey}

We summarise the methods most often used in practice, grouped by their dominant
analytical stance, then compare them along ten dimensions.

\subsection{System- and software-centric elicitation}

\paragraph{STRIDE.} Introduced at Microsoft by Kohnfelder and
Garg~\cite{kohnfelder1999threats} and later embedded in the Security
Development Lifecycle~\cite{microsoft-sdl,shostak2014threat}, STRIDE is a
mnemonic taxonomy: \emph{Spoofing}, \emph{Tampering}, \emph{Repudiation},
\emph{Information disclosure}, \emph{Denial of service}, and \emph{Elevation of
privilege}. Each category is the violation of a desired property
(authentication, integrity, non-repudiation, confidentiality, availability,
authorisation). The method is typically applied over a data-flow diagram (DFD),
in the \emph{STRIDE-per-element} or \emph{STRIDE-per-interaction} variants. Its
strengths are completeness of coverage for conventional components and low
adoption cost. Its weaknesses are well documented: threat explosion on
non-trivial DFDs, dependence on diagram quality, no native risk ranking (the
companion DREAD rating was deprecated for subjectivity), and a property
taxonomy that says nothing about \emph{who} the actor is or whether the action
was authorised.

\paragraph{Trike.} Trike~\cite{trike2005} is a risk-based, requirements-driven
method built around an actor-asset-action matrix and an explicit acceptable-risk
model. It auto-derives threats (limited to elevation of privilege and denial of
service) from a CRUD-style permission matrix. It introduces useful rigour
around who is permitted to do what, but it is tooling-dependent, has seen
limited maintenance, and remains application-scoped.

\paragraph{LINDDUN.} LINDDUN~\cite{deng2011linddun} mirrors STRIDE's structure
for the privacy domain: \emph{Linkability}, \emph{Identifiability},
\emph{Non-repudiation}, \emph{Detectability}, \emph{Disclosure of information},
\emph{Unawareness}, and \emph{Non-compliance}. Its inclusion here makes a
general point that motivates TRACE: a threat model is strongest when its
taxonomy matches the risk domain. In LINDDUN it is the taxonomy, not the
diagramming, that does the analytical work.

\subsection{Risk- and attacker-centric process methods}

\paragraph{PASTA.} The Process for Attack Simulation and Threat Analysis
\cite{ucedavelez2015risk} is a seven-stage, risk-centric methodology:
(1)~define business objectives, (2)~define the technical scope, (3)~decompose
the application, (4)~analyse threats, (5)~analyse vulnerabilities and
weaknesses, (6)~model attacks, and (7)~analyse risk and impact. PASTA's
strength is its explicit linkage of technical threats to business impact and
its attacker simulation step. Its cost is weight: it is heavyweight,
expertise-intensive, and oriented around an application or product rather than
a cross-organisational control fabric.

\paragraph{Attack trees.} Popularised by Schneier~\cite{schneier1999attack},
attack trees are an AND/OR decomposition of an attacker goal into sub-goals and
leaf actions, optionally annotated with cost, feasibility, or probability. They
are a depth instrument rather than a discovery instrument: they answer ``how
could goal $G$ be achieved?'' but presuppose that the analyst has already
identified $G$. TRACE uses them in exactly this complementary role.

\subsection{Organisation- and asset-centric methods}

\paragraph{OCTAVE / OCTAVE Allegro.} OCTAVE
\cite{alberts2002octave} and its streamlined successor OCTAVE
Allegro~\cite{caralli2007allegro} are asset- and organisation-centric risk
assessment frameworks from the SEI. They start from organisationally critical
information assets and their containers, and they explicitly engage operational
staff rather than only architects. This is the closest pre-existing method to
TRACE's organisational pillar. Its limitations for the present setting are that
it is information-asset oriented (less suited to value, key, and authority
assets), workshop-heavy, and not designed to integrate with technical attack
decomposition or distributed-authority analysis.

\paragraph{VAST.} The Visual, Agile, and Simple Threat methodology, embodied in
the ThreatModeler tool~\cite{threatmodeler-vast}, targets enterprise
scalability by distinguishing application threat models (process-flow diagrams)
from operational threat models and integrating with CI/CD. Its emphasis on
scale and automation is relevant, but it is a commercial tooling approach
rather than an open methodology, and it does not introduce invariants,
collusion, or distributed-authority primitives.

\paragraph{Hybrid and human-factor methods.} The SEI's Hybrid Threat Modeling
Method (hTMM)~\cite{mead2018htmm} combines Security
Cards~\cite{denning2013securitycards} and Persona non
Grata~\cite{clelandhuang2014png} with SQUARE to balance coverage against false
positives. Security Cards and PnG specifically push analysts to reason about
adversary archetypes, motivations, and human impact, a dimension absent from
STRIDE. These remain brainstorming aids rather than full structured pipelines.

\subsection{Adversary-behaviour knowledge bases}

\paragraph{Cyber Kill Chain and MITRE ATT\&CK.} The Lockheed Martin Cyber Kill
Chain~\cite{hutchins2011killchain} models an intrusion as an ordered sequence
of phases (reconnaissance through actions on objectives). MITRE
ATT\&CK~\cite{strom2018attack} is a far richer, empirically grounded knowledge
base of adversary tactics, techniques, and procedures (TTPs). Neither is a
design-time threat \emph{modelling} method; they are post-design,
threat-informed-defence and detection-engineering instruments. They describe
\emph{how} known adversaries operate against deployed systems, not \emph{what}
a particular system must protect or which of its assumptions are incoherent.

\subsection{Standards context}

For completeness, NIST SP 800-30~\cite{nist80030} frames risk assessment,
SP 800-154~\cite{nist800154} proposes data-centric system threat modelling,
SP 800-207~\cite{nist800207} specifies zero trust architecture, and OWASP
\cite{owasp-tm} captures the general four-question loop (what are we building,
what can go wrong, what will we do, did we do a good job). These are reference
frames rather than competing methodologies, and TRACE positions itself relative
to zero trust in particular (Section~\ref{sec:systems}).

\subsection{Comparison}

Table~\ref{tab:comparison} compares the surveyed methods across ten dimensions
chosen to expose the distributed, cloud-first setting. The pattern is
consistent: each method is strong in its origin domain and structurally silent
on the others. No existing method simultaneously treats invariants, distributed
authority, collusion, control-plane/supply-chain edges, \emph{and}
human/operational risk as first-class objects.

\begin{table}[t]
\centering
\caption{Threat modelling and adversary-analysis methods across ten dimensions.
\yes{}~= first-class / native support; \prt{}~= partial or ad hoc; \no{}~= not
addressed. Dimensions: \textbf{Pers.}~primary perspective (Sys=system/asset,
Att=attacker, Org=organisation, Adv=adversary-behaviour); \textbf{Rank}~native
risk ranking; \textbf{Inv}~critical invariants as objects; \textbf{Auth}~distributed/
privileged authority; \textbf{Coll}~collusion \& coordination; \textbf{Supply}~CI/CD
\& supply-chain edges; \textbf{Cloud}~cloud/SaaS control-plane \& zero-trust fit;
\textbf{Human}~human/operational failure; \textbf{Trace}~evidence traceability;
\textbf{AI}~human-AI co-working by design.}
\label{tab:comparison}
\footnotesize
\setlength{\tabcolsep}{3pt}
\renewcommand{\arraystretch}{1.25}
\begin{tabular}{l c c c c c c c c c c}
\toprule
\textbf{Method} & \textbf{Pers.} & \textbf{Rank} & \textbf{Inv} & \textbf{Auth} & \textbf{Coll} & \textbf{Supply} & \textbf{Cloud} & \textbf{Human} & \textbf{Trace} & \textbf{AI}\\
\midrule
STRIDE              & Sys & \no   & \no   & \no   & \no & \prt & \prt & \no   & \prt & \no \\
Trike               & Sys & \yes  & \no   & \prt & \no & \no   & \no   & \no   & \prt & \no \\
LINDDUN             & Sys & \prt & \no   & \no   & \no & \no   & \no   & \no   & \prt & \no \\
PASTA               & Att & \yes  & \prt & \prt & \no & \prt & \prt & \prt & \prt & \no \\
Attack trees        & Att & \prt & \no   & \prt & \prt & \prt & \prt & \prt & \no  & \no \\
OCTAVE (Allegro)    & Org & \yes  & \prt & \prt & \no & \no   & \no   & \yes  & \prt & \no \\
VAST                & Sys & \prt & \no   & \no   & \no & \prt & \prt & \no   & \prt & \no \\
hTMM / PnG          & Adv & \prt & \no   & \prt & \prt & \no  & \no   & \prt & \no  & \no \\
Kill Chain / ATT\&CK & Adv & \prt & \no   & \no   & \no & \prt & \prt & \no   & \prt & \no \\
\midrule
\textbf{TRACE}      & \textbf{Sys+Org+Att} & \yes & \yes & \yes & \yes & \yes & \yes & \yes & \yes & \yes \\
\bottomrule
\end{tabular}
\end{table}

\section{Why Existing Methods Fall Short in Distributed, Cloud-First Settings}
\label{sec:gaps}

The shortfalls below are not defects of the surveyed methods within their design
scope; they are consequences of applying perimeter-and-product assumptions to an
environment that no longer satisfies them.

\subsection{The perimeter has dissolved into edges}

STRIDE-over-DFD and similar methods presume that trust is granted at a boundary
between trusted and untrusted zones. In a cloud-first organisation there is no
single boundary: there are many \emph{edges} where value, control, data, or
authority crosses between an identity provider and a cloud account, a CI job and
a production deployment, a vendor and an internal system, a signer and an
approval. Zero trust~\cite{nist800207} exists because each such crossing must be
evaluated per request. Existing methods can annotate a boundary on a diagram,
but they do not make the \emph{edge} the unit of analysis, nor do they ask the
zero-trust question (``what if the source of this request is already
compromised?'') systematically at every crossing.

\subsection{Authorisation is modelled; authority is not}

STRIDE's elevation-of-privilege category, and Trike's permission matrix, ask
whether an actor can perform an action it is not permitted to perform. They are
largely silent on the actor who \emph{is} permitted and acts harmfully: the
malicious or compromised administrator, deployer, signer, or governance
delegate. In distributed organisations this is the dominant risk: a privileged
role operating entirely within its formal permissions can drain value, ship a
backdoored release, or capture governance. The analyst needs to ask not only
``can this action succeed?'' but ``what can this actor cause?''. Authority, as
distinct from authorisation, is not a first-class object in any surveyed method.

\subsection{Collusion and coordination are out of scope}

No mainstream threat modelling method models combinations of actors. Yet
multisig quorums, governance committees, validator sets, vendor relationships,
and approval ceremonies all fail under coordinated rather than single-actor
behaviour. The relevant questions (can $k$ of $n$ signers collude? are two
``independent'' vendors actually independent? is a committee quorum capturable
slowly?) are the subject of distributed-systems theory (Byzantine fault
tolerance~\cite{lamport1982byzantine}, accountable
consensus~\cite{civit2023abc}) but absent from the threat-modelling toolbox.
Treating each actor in isolation systematically underestimates risk in any
system whose safety rests on a threshold or separation-of-duties assumption.

\subsection{The supply chain and control plane are the attack surface}

The highest-impact recent failures targeted the build and deployment control
plane rather than the running application: a compromised software update
pipeline~\cite{mehdi2021solarwinds}, package-registry and dependency
compromise, CI secrets exfiltration, and IAM misconfiguration that converts a
cloud mistake into production control. STRIDE and PASTA can be stretched to
cover these, but they do not natively model CI/CD-to-deployment authority,
artifact integrity, or third-party control planes as core objects. ATT\&CK
catalogues such techniques descriptively but offers no design-time model of the
specific organisation's build-and-release authority graph.

\subsection{Human and operational reality is treated as background}

OCTAVE and the human-factor methods (Security Cards, PnG) come closest: they
engage people, but they do not connect human and procedural failure back to
technical assets and invariants in one model.
Meanwhile decentralized, remote-first organisations frequently concentrate
catastrophic authority in a few highly privileged individuals. A laptop
compromise, an unclear approval ceremony, a rushed incident response, a
deceptive signing interface~\cite{fbi2025bybit}, or an account-recovery path can
be as decisive as any code vulnerability. A method that models the software but
not the operational behaviour around it gives a false sense of coverage.

\subsection{Invariants are never named}

Underlying all of the above: existing methods enumerate \emph{threats}
(violations of generic properties) but rarely force the analyst to state the
specific \emph{invariants} the system must preserve: solvency, segregation of
duties, approval integrity, bounded authority, deployment integrity, recovery
ability. Without explicit invariants, threat ranking has no anchor: severity
becomes a function of how technical or recent a component looks rather than what
it would break. Invariants are the missing semantic layer between assets and
threats.

\subsection{Methods predate human-AI co-working}

Finally, every surveyed method was specified before large language models could
plausibly ingest a heterogeneous source corpus and draft model objects, STRIDE
candidates, and attack-tree branches. Used naively, an LLM amplifies the
failure modes above: it produces generic perimeter-era threat lists,
plausible-but-unsupported architecture assumptions, and attack trees that are
structurally elegant yet operationally implausible. None of the existing
methods provide a discipline
for incorporating such assistance while preserving expert judgement and
evidentiary traceability.

\section{TRACE}
\label{sec:trace}

TRACE is a threat modelling methodology that turns heterogeneous source material
into a structured, evidence-linked model of assets, roles, invariants, trust
boundaries, value flows, authority paths, and failure paths, and then moves from
that model to ranked threats, attack trees, collusion analysis, and a
prioritised mitigation roadmap. It is evidence-driven: every material threat
should be traceable back to a source, a model object, an assumption, a boundary,
or an attack path.

It was developed at Oak Security through Web3 security work, where high-value
assets, distributed authority, governance, off-chain infrastructure, and human
operations are tightly coupled and therefore expose the gaps of
Section~\ref{sec:gaps} in concentrated form. The method is stack-agnostic and
applies to any organisation with fragmented control paths, cloud and SaaS
dependency chains, critical human approvals, externally operated systems, and no
clear perimeter.

\subsection{Core model}

TRACE makes five object types first-class. The acronym names them:

\begin{description}[leftmargin=1.6em,itemsep=3pt]
  \item[\textbf{T}hreat actors.] Actors with capability, incentive, or authority
  to affect the target: external attackers, insiders, vendors, contractors,
  service providers, administrators, delegates, compromised users, economic
  adversaries, governance participants.
  \item[\textbf{R}oles.] Privileged or operational positions inside the target:
  founders, executives, signers, maintainers, deployers, administrators,
  operators, responders, service owners. Roles localise \emph{authority}.
  \item[\textbf{A}ssets.] Value, control, data, authority, or continuity that
  must be protected: funds, keys, credentials, production control, customer
  data, uptime, brand trust, source code, governance power.
  \item[\textbf{C}ritical invariants.] Properties that must remain true:
  segregation of duties, approval integrity, data integrity, financial
  correctness/solvency, bounded authority, deployment integrity, recovery
  ability. Invariants are the anchor for ranking.
  \item[\textbf{E}dges.] Places where trust, value, data, or control crosses
  domains: trust boundaries, value flows, signer paths, API boundaries, admin
  paths, CI/CD-to-deployment, identity-provider-to-cloud. Edges are the unit of
  zero-trust analysis.
\end{description}

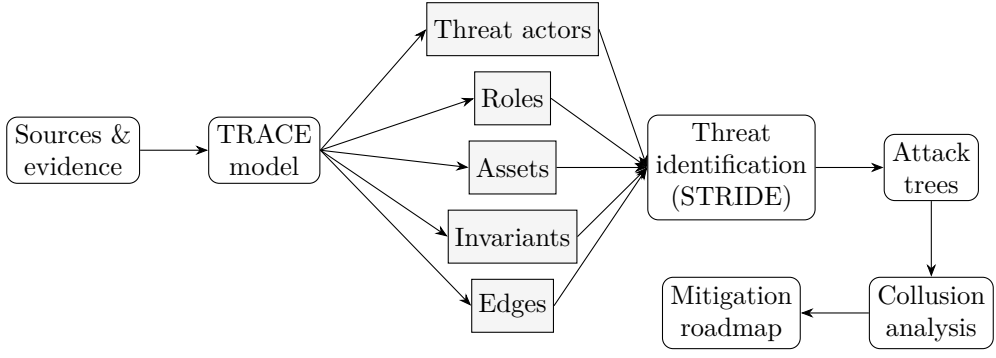
\begin{figure}[t]
\centering
\begin{tikzpicture}[
  font=\small,
  node distance=6mm and 9mm,
  box/.style={draw, rounded corners, align=center, minimum height=8mm, inner sep=3pt},
  obj/.style={draw, align=center, minimum height=7mm, inner sep=2.5pt, fill=black!4},
  >={Stealth[]}
]
  \node[box] (src) {Sources \&\\ evidence};
  \node[box, right=of src] (model) {TRACE\\ model};
  \node[obj, right=14mm of model, yshift=16mm] (a) {Threat actors};
  \node[obj, below=2mm of a] (r) {Roles};
  \node[obj, below=2mm of r] (as) {Assets};
  \node[obj, below=2mm of as] (i) {Invariants};
  \node[obj, below=2mm of i] (e) {Edges};
  \node[box, right=12mm of as] (thr) {Threat\\ identification\\ (STRIDE)};
  \node[box, right=of thr] (tree) {Attack\\ trees};
  \node[box, below=10mm of tree] (col) {Collusion\\ analysis};
  \node[box, left=of col] (road) {Mitigation\\ roadmap};

  \draw[->] (src) -- (model);
  \draw[->] (model.east) -- (a.west);
  \draw[->] (model.east) -- (r.west);
  \draw[->] (model.east) -- (as.west);
  \draw[->] (model.east) -- (i.west);
  \draw[->] (model.east) -- (e.west);
  \draw[->] (a.east) -- (thr.west);
  \draw[->] (r.east) -- (thr.west);
  \draw[->] (as.east) -- (thr.west);
  \draw[->] (i.east) -- (thr.west);
  \draw[->] (e.east) -- (thr.west);
  \draw[->] (thr) -- (tree);
  \draw[->] (tree) -- (col);
  \draw[->] (col) -- (road);
\end{tikzpicture}
\caption{TRACE pipeline: heterogeneous evidence is condensed into a structured
model of five first-class object types, which drives STRIDE elicitation, attack
trees on the top-ranked threats, an optional collusion pass, and a mitigation
roadmap. Every downstream artifact links back to a model object and to source
evidence.}
\label{fig:model}
\end{figure}

TRACE separates \emph{authority} (located on roles, exercised across
edges) from \emph{authorisation} (a per-action permission check), and it treats
\emph{invariants} as the semantic bridge from assets to threats. These two
design choices directly address the gaps of Sections~\ref{sec:gaps} ($\S$3.2)
and ($\S$3.6). Figure~\ref{fig:model} shows the flow from evidence to roadmap.

\subsection{The three pillars}

The same core method is applied at three layers, varying the input material and
model emphasis. Failure routinely crosses layers, so a complete assessment uses
all three.

\paragraph{TRACE for Protocols (design stage).} A protocol is any formal or
semi-formal rule system governing value, authority, access, coordination, or
critical behaviour: a smart-contract system, validator network, or governance
process in Web3; an approval protocol, access policy, AI-agent control policy,
or financial process in a broader organisation. Inputs are specifications, white
papers, mechanism and policy descriptions, economic models, governance
proposals, and code where available. The model emphasises assets, invariants,
privileged protocol roles, governance and approval authority, value/authority
flows, external dependencies, and collusion/quorum/delegation/consensus
assumptions. Outputs include a design-level threat model, an invariant map, a
governance and privileged-role risk analysis, a STRIDE catalogue by component or
flow, attack trees for prominent threats, and a collusion/consensus-surface
analysis where relevant.

\paragraph{TRACE for Systems (architecture and infrastructure).}
\label{sec:systems}
Inputs are system diagrams, architecture and deployment documents, cloud account
and IAM structure, identity-provider and SaaS configuration, CI/CD flows,
infrastructure-as-code, dependency and package-registry information, and
logging/alerting/incident runbooks. The model emphasises trust boundaries
between internal, cloud, SaaS, vendor, and public systems; build/deploy/release
authority; infrastructure control planes; privileged credential and key paths;
frontend/API/data integrity; operational automation; and dependency/supply-chain
exposure. This pillar composes directly with zero trust
architecture~\cite{nist800207,cisa-ztmm}: each TRACE edge becomes a zero-trust
question. Who or what is crossing, what identity is asserted, what device or
workload, what resource, what action, what context changes the risk, and what
happens if the source is already compromised? Outputs include a trust-boundary
inventory, a zero-trust gap map for critical access paths, a CI/CD and
supply-chain risk map, and infrastructure attack trees.

\paragraph{TRACE for Organisations (operational security).} Inputs are discovery
workshops, interviews, access reviews, device and account inventories, custody
and signing and approval procedures, incident-response plans, vendor and
contractor relationships, travel and physical-security assumptions, and observed
team practice. The model emphasises human authority over assets and invariants,
privileged people and groups and vendors, account-recovery and access-restoration
paths, approval workflows and exception paths, incident coordination, and
social/procedural failure modes. Outputs include an operational threat model, a
human-and-process risk register, custody/approval risk analysis, an incident
readiness assessment, a vendor and access-risk map, and a 30/60/90-day
operational hardening roadmap.

\subsection{Workflow and approval gates}

TRACE is deliberately sequential. Each phase produces a reviewable artifact that
becomes input to the next, and explicit human approval gates prevent weak
assumptions from compounding into threat rankings, attack trees, or
recommendations.

\begin{enumerate}[leftmargin=1.6em,itemsep=2pt]
  \item \textbf{Scope and source inventory.} Define the assessment boundary and
  lifecycle stage; collect evidence; record assumptions, exclusions, and missing
  sources that create uncertainty. \emph{Gate:} a senior reviewer approves
  scope and known gaps.
  \item \textbf{Ingest sources.} Extract candidate components, actors, roles,
  assets, invariants, dependencies, boundaries, and flows.
  \item \textbf{Construct the TRACE model.} Build the five object sets; record
  evidence and assumptions; mark any object not tied to a source as an inferred
  assumption. \emph{Gate:} a senior reviewer approves and complements the model
  before threat expansion.
  \item \textbf{STRIDE identification and ranking.} Apply STRIDE across
  components, flows, roles, and edges; rank by impact on assets and invariants,
  feasibility, incentive compatibility, existing mitigations, blast radius,
  time sensitivity, and model confidence. \emph{Gate:} reviewer approves the
  threat set and ranking.
  \item \textbf{Build attack trees.} Decompose the top-ranked threats from goal
  to enabling conditions; map each root to a ranked threat and each leaf to an
  enabling assumption or concrete fact and to a candidate mitigation.
  \emph{Gate:} attack trees approved for plausibility and missing branches.
  \item \textbf{Inspect collusion and coordination surfaces.} Where actor groups
  can coordinate (committees, multisigs, validators, vendors, operators),
  examine actor combinations, quorum/threshold assumptions, incentive alignment,
  governance-capture paths, and operational dependencies between nominally
  independent parties. \emph{Gate:} reviewer validates which coordination paths
  are credible.
  \item \textbf{Produce roadmap and report.} Executive summary, model, ranked
  threats, attack trees, collusion findings, mitigation roadmap, and labelled
  open questions. \emph{Gate:} recommendations approved for severity,
  feasibility, and sequencing.
\end{enumerate}

\section{Human-AI Co-Working by Design}
\label{sec:ai}

TRACE is built for human-AI co-working, not for autonomous threat modelling.
Large language models are effective accelerators for reading large, heterogeneous
source sets, extracting candidate model objects, checking coverage, proposing
STRIDE candidates, drafting attack-tree branches, and maintaining traceability
between evidence, model objects, threats, and recommendations. They are
unreliable precisely where judgement matters most: defining the critical
invariant, calibrating severity, and deciding which collusion paths are
credible.

The methodology therefore treats AI output as \emph{candidate analysis} until
reviewed, and enforces an evidence-and-traceability discipline:

\begin{itemize}[leftmargin=1.4em,itemsep=1.5pt]
  \item Model objects link back to sources, or are explicitly marked inferred.
  \item STRIDE threats link to components, flows, roles, or edges.
  \item Attack-tree roots link to ranked threats; leaves link to enabling
  assumptions or concrete facts.
  \item Recommendations link to threats, attack-tree leaves, assets, invariants,
  or trust boundaries.
\end{itemize}

Any claim that cannot be traced is removed, rewritten as an assumption, or
recorded as an open question. The sequential gates of Section~\ref{sec:trace}
exist to catch the characteristic AI failure modes: overconfident
claims from incomplete sources, plausible-but-unsupported architecture
assumptions, generic perimeter-era threat lists, conflation of formal permission
with operational safety, omission of human/vendor/signer realities, and
structurally elegant but unrealistic attack trees. AI output is never used
directly for final invariant definitions, final severity ratings, collusion or
governance-capture conclusions, exploitability claims, or final
recommendations.

\section{Discussion}

\paragraph{Relationship to existing methods.} TRACE does not discard the toolbox;
it re-bases it. STRIDE remains the elicitation engine (step~4), attack trees the
depth instrument (step~5), and PASTA/OWASP inform the broader risk loop. Zero
trust~\cite{nist800207,cisa-ztmm} supplies the per-edge evaluation discipline,
and Byzantine/accountable-consensus theory~\cite{lamport1982byzantine,civit2023abc}
informs the collusion pass. TRACE's contribution is the set of first-class
primitives (invariants, authority-bearing roles, edges, and collusion surfaces)
together with the evidence-linked, gated workflow that connects them, which is
what lets the older instruments operate over a perimeter-free, multi-party
control fabric.

\paragraph{Where TRACE fits.} TRACE is a modelling layer, not a replacement for
implementation audits, compliance assessments, or zero-trust implementation. It
is most useful before an audit (to surface the design and operational
assumptions reviewers should focus on), before launch (to locate where risk is
concentrated: code, infrastructure, governance, key management, identity,
vendors, or operations), and during live operation (to revisit assumptions as
teams, vendors, and integrations evolve).

\paragraph{Limitations.} TRACE is, at present, a methodology validated through
professional practice rather than a controlled empirical study. The first-class
object set adds modelling overhead relative to a quick STRIDE pass, and the
collusion and organisational pillars depend on access to people and procedures
that may be unavailable or misrepresented. The quality of the model is bounded
by source quality; the inferred-assumption mechanism mitigates but does not
eliminate this. Severity ranking, while anchored to invariants, still rests on
expert judgement and is not fully reproducible across analysts. The human-AI
discipline reduces but cannot fully exclude automation bias at the review gates.

\paragraph{Future work.} Three directions follow. First, empirical evaluation:
inter-analyst agreement on invariants and rankings, and comparison of TRACE
against STRIDE-only baselines on coverage of authorised-path and collusion
threats. Second, formalisation: expressing invariants and edges in a
machine-checkable schema so that traceability constraints can be validated
automatically. Third, tighter coupling to accountable-consensus and
mechanism-design results so that collusion thresholds can be reasoned about
quantitatively rather than qualitatively.

\section{Conclusion}

Existing threat modelling frameworks remain valuable within the assumptions for
which they were designed: a perimeter, a single owning organisation, and the
equation of authorisation with safety. Distributed, cloud-first organisations
satisfy none of these. The dominant failures in such environments (authorised
actors behaving harmfully, collusion across nominally independent parties,
control-plane and supply-chain compromise, and operational mishandling of
high-value actions) fall in the structural blind spots of the established
methods. TRACE addresses these by promoting threat actors, roles, assets,
critical invariants, and edges to first-class, evidence-linked objects;
separating authority from authorisation; making invariants the anchor for
ranking; spanning protocols, systems, and organisations in one model; and
embedding a gated human-AI workflow that accelerates coverage without ceding
judgement. The framework emerged from Web3 security practice, where these
pressures are most concentrated, but the problem it addresses is now general.
The methodology specification is released openly under CC BY 4.0.

\section*{Acknowledgements}

TRACE was developed through security engagements at Oak Security. We thank the
Oak Security team and the protocol, infrastructure, and operational teams whose
engagements shaped the methodology.

\bibliographystyle{plainnat}
\bibliography{refs}

\end{document}